\documentclass{PoS}
\usepackage{amsmath}
\newcommand{\Section}[1]{\vspace{-0.15cm}\section{#1}\vspace{-0.15cm}}
\newcommand{\Subsection}[1]{\vspace{-0.15cm}\subsection{#1}\vspace{-0.15cm}}

\def \Qs {Q_{\rm s}}
\def \d {{\rm d}}

\title{Thermalization in collisions of large nuclei at high energies}

\ShortTitle{Thermalization in collisions of large nuclei at high energies}

\author{\speaker{Aleksi Kurkela }\\
        McGill University\\
        E-mail: \email{aleksi.kurkela@mcgill.ca}}

\abstract{
Hydrodynamical analysis of experimental data of ultrarelativistic heavy ion collisions seems to indicate that the hot QCD matter created in the collisions thermalizes very quickly. Theoretically, we have no idea why this should be true. In this proceeding, I will describe how the thermalization takes place in the most theoretically clean limit -- that of large nuclei at asymptotically high energy per nucleon, where the system is described by weak-coupling QCD. In this limit, plasma instabilities dominate the dynamics from immediately after the collision until well after the plasma becomes nearly in equilibrium at time $t \sim \alpha_s^{-5/2} \Qs^{-1}$.
}

\FullConference{Xth Quark Confinement and the Hadron Spectrum,\\
		October 8-12, 2012\\
		TUM Campus Garching, Munich, Germany}

\begin{document}

\Section{Introduction}
\enlargethispage*{\baselineskip}
The hydrodynamical description of heavy ion collisions at RHIC and at LHC has led to a tremendous
phenomenological success. However, the hydrodynamical treatment can be
justified only if the matter created in the collision is near local thermal equilibrium (or more precisely close to local isotropy).
At the initial stages of the collision, this condition is clearly violated and it is an open theoretical question,
how quickly --- and how --- the matter approaches equilibrium. It is crucial to understand this prethermal evolution as our ignorance of it constitutes
one of the largest systematic uncertainties in analysis of the heavy-ion collisions.

In \cite{KM1}, we have addressed this problem in the most
theoretically clean limit --- that of large nuclei at asymptotically high energy per nucleon,
where the system is described by weak-coupling QCD. When the typical energy scale right 
after the collisions is large $\Qs \gg \Lambda_{\rm QCD}$, the renormalized strong coupling constant becomes small $\alpha(\Qs)\ll 1$.
In this limit, the initial
condition for the collision is understood in the color glass condensate (CGC) framework \cite{CGC}. 
 In \cite{KM1}, we have identified the most important physical processes, in the $\alpha\ll 1$ limit, 
 that drive the evolution from the initial strong and anisotropic CGC fields to the thermal state
 in the longitudinally expanding geometry of a heavy-ion collision. Our solution resembles the original ``Bottom-Up'' 
 thermalization \cite{BMSS}, with the difference that it takes into 
 account the physics of plasma instabilities \cite{instab}.
 
The evolution proceeds in three
stages.
The first stage ($1 \ll \Qs \tau \ll \alpha^{-\frac{8}{7}}$) is characterized by strong fields, or equivalently high occupancies ($f\gg 1$).
It
is a competition between the longitudinal expansion that
drives the system towards larger anisotropies ($ \langle | p_\perp |\rangle \gg \langle |p_z| \rangle  \equiv \alpha^d \langle | p_\perp |\rangle$) and weaker fields, 
and momentum broadening due to interactions that works towards isotropizing the fields.
The result of this competition is that the anisotropy increases and the occupancies 
decrease as a function of time.

During the second stage $\alpha^{-\frac{8}{7}} \ll \Qs \tau \ll \alpha^{-\frac{12}{5}}$,
the system is highly anisotropic but the typical modes are now under-occupied, $f\ll 1$. 
The cross-over from high to low occupancy changes qualitatively the system's behavior:
inelastic scattering begins to increase particle number. In particular, soft
inelastic emissions create a bath soft gluons that eventually becomes nearly thermal. 
During this stage, the soft protothermal bath does not dominate anything; the primordial hard 
particles carry the most energy, inflict most screening, are more numerous, and
cause the most scattering.

Eventually the soft protothermal bath will, however, start to dominate the physics, 
and in particular the momentum broadening experienced by the hard particles. 
This is the third stage  $\alpha^{-12/5} \ll \Qs \tau \ll \alpha^{-5/2}$. 
During this stage the hard anisotropic particles still carry most of the energy density of the system
(and hence dominate $T_{\mu\nu}$, relevant for hydrodymamics), but they effectively decouple from each other as the
dominant interaction is with the bath of soft particles; the hard particles
can be seen as few but highly energetic ``jets'' propagating in a nearly thermal medium. 
The thermalization then proceeds through quenching of these jets: interaction
with the protothermal medium leads to hard collinear splitting and therefore to 
radiative energy loss. Once the hard modes have had time to lose all their energy
to the medium (by the time $Q_s \tau \sim \alpha^{-5/2}),$ all that is left is the nearly thermal bath of soft modes, and 
the system has essentially thermalized. 

Throughout the evolution, the dominant interaction between the hard particles, and between the 
hard particles and the protothermal bath, takes place via plasma (or Chromo-Weibel) unstable modes:
in an anisotropic system, a set of long-wavelength chromo-magnetic fields is perturbatively unstable and the unstable modes undergo an exponential growth until they become non-perturbative (with $f(k_{\rm inst})\sim 1/\alpha $)
and saturate. When hard particles propagate through a stochastic background of these saturated $B$-fields, 
they exchange momentum with the $B$-fields due to Lorentz force. As the $B$-fields are at long scales incoherent,
this momentum transfer is diffusive and is described by a (time dependent) momentum
diffusion coefficient $\hat{q}_{\rm inst}\sim {\rm d} p^2/{\rm d} t$. This $\hat{q}$ turns out to be
larger than that arising from ordinary elastic scattering at all stages of thermalization.

During the first and second stages, the most
important instabilities are those due to the hard particles. The soft protothermal 
bath is also sightly anisotropic (because of expansion and anisotropic ``rain'' of soft particles arising from splitting 
of the anisotropic hard particles), and during the third stage it is the magnetic fields
that became unstable because of the protothermal bath that dominate the 
momentum transfer. 

Here, we discuss these three stages in slightly more detail, but with much less detail
than \cite{KM1}, concentrating only on the dominant scales.

\Section{Initial condition in heavy ion collisions: Color-Glass-Condensate}
In the weak coupling limit, the very early dynamics are well understood in the 
color glass condensate (CGC) framework \cite{CGC}.
It indicates that at time $\Qs\tau \sim 1$, the system consists of intense,
nearly boost invariant gluon fields, with a coherence length $l_{\rm coh}\sim Q_s^{-1}$
in the $xy$-plane transverse to the beam axis, and a much longer coherence length in the $z$-direction
along the beam axis, and that the energy density of the system is $e (\tau \sim Q_s^{-1})\sim Q_s^{4}/\alpha$ .

\begin{figure}
\vspace{-0.4cm}
\begin{center}
\includegraphics[width=0.175\textwidth]{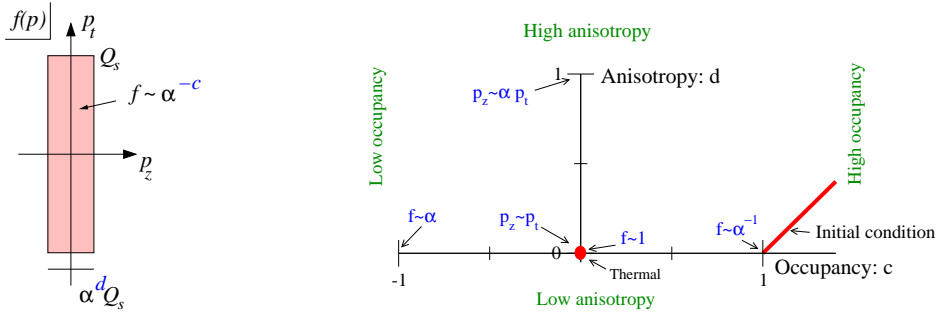}
\hspace{2cm}
\includegraphics[width=0.500\textwidth]{cdplane0c.eps}
\caption{
(Left) Graphical representation of the descriptors of an anisotropic single-scale system. (Right) The CGC initial condition lies of a $d = c-1$ line on $cd$-plane.
\vspace{-0.8cm}
}
\label{fig1}
\end{center}
\end{figure}

At times $\Qs\tau>1$, the fields have lost phase coherence, and can be described in terms of particle degrees of freedom. The corresponding distribution
function of gluons can be parametrized by%
\footnote{
The $\theta$-functions should not be interpreted as sharp step functions but rather as smooth 
momentum cutoffs so that the actual form of the distribution is such that high momentum particles with $p>Q_s$ do not dominate anything, \emph{i.e.} $f(p > Q_s)<(Q_s/p)^4$.
This ansatz for the distribution is accurate enough for parametric estimates (of powers of $\alpha$), for full numerical 
description one needs to consider the $\mathcal{O}(1)$ details of the distribution function. 
Also, it assumes that the physics of the system is dominated by a single scale only, which 
need not to be the case. However, multi-scale systems can be constructed by superimposing
several distribution functions of this kind, as we will do in Section \ref{sec:bottomup}. The functional (power-law) form of $f$
is described in detail in $\cite{KM1}$, here we limit ourselves to
this simpler description for clarity.
}
\begin{equation}
f(\vec{p}) \sim \alpha^{-c} \theta( Q_s - |p|)\theta( \alpha^{d} Q_s- p_z ), \quad d > 0
\label{dist}
\end{equation}
where $c$ describes the typical occupation number, and $d$ parametrizes the anisotropy. In terms of these descriptors, the energy density condition implies that
on the $cd$-plane, at times $\Qs \tau \gtrsim 1$, the system lies on a $d = c-1$ line, as shown in
Fig.~\ref{fig1}. We take this as our initial condition. 

Where precisely on the line the initial condition sits is a matter of taste. Both, our description and that within CGC framework \cite{glasmamelt}, displays that the system evolves to $\mathcal{O}(1)$ anisotropy
in a time $\tau \sim Q_s^{-1}\times(\textrm{logs of }\alpha)$. Therefore, at times $\tau Q_s \gg 1$, but less than any negative power of $\alpha$, the system has
$c=1$ and $d = 0$.%

\Section{Evolution to small occupancies $1 \ll Q_s \tau \ll \alpha^{-12/5} $}
As long as $c=1$ and $d=0$, the dynamics of the system are 
fully non-perturbative, and amendable only to non-perturbative (yet classical) simulations \cite{glasmamelt,clas}.
However, the system is only marginally non-perturbative; deviation
from this point by any positive power of $\alpha$ ($c<1$ or $d>0$) renders parts of the system perturbative. 
And indeed, the interactions and the expansion both work toward driving the system away from this point.

Once the system is away from the non-perturbative point, several scale separations emerge. In particular, 
the kinetic mean free path of particles with $p\sim Q_s$ becomes much larger than their
thermal wavelength, $l_{free} \sim \alpha^{-2-d+2c} Q_s^{-1}\gg Q_s^{-1}$,
so that their evolution and mutual interactions can be described in an effective kinetic theory \cite{kinetic}. The screening
scale $(m^2 \sim \int {\rm d} ^3p f(p)/p \sim \alpha^{d-c} Q_s^2)$ becomes parametrically longer than
the wavelength of the typical hard particles. The interaction between the modes at scales $m$ and $\Qs$  
is still non-perturbative, but between modes at scale $m$ perturbative. The non-perturbative interaction between $m$ and $\Qs$ scales can be
resummed by describing the modes in the screening scale by 
collisionless Vlasov equations, or equivalently in the hard-loop (HL) effective theory. 
The non-equilibrium system also has an infrared scale, analogous to the magnetic scale ($\alpha T$) of the thermal ensemble, 
characterized by non-perturbative interactions between modes at the infrared scale. Fortunately, this scale does
not dominate anything.%
\footnote{In \cite{BE1} it has been argued that the infrared scale might dominate particle number and contain 
a Bose-Einstein condensate. For further discussion in this topic see \cite{KM3}.}

\begin{figure}
\begin{center}
\includegraphics[width=0.54\textwidth]{cdplane4.eps}
\hspace{0.3cm}
\includegraphics[width=0.37\textwidth]{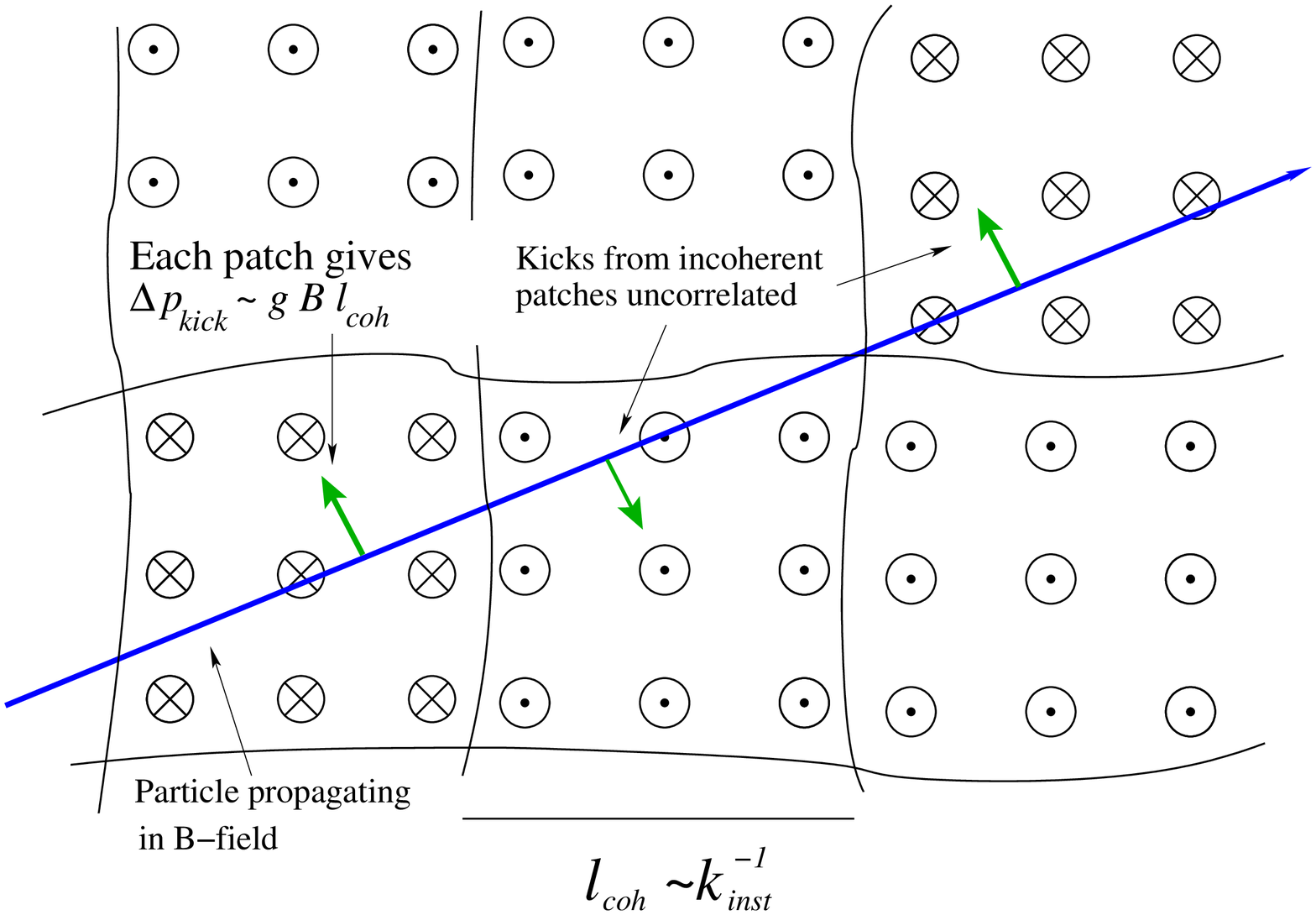}
\end{center}
\vspace{-0.5cm}
\caption{(Left) Solutions to the evolution in $cd$-plane. In absence of interactions the system follows ``Free streaming'' line. Including perturbative scattering but
neglecting instabilities leads to the attractor labeled ``elastic'' from \cite{BMSS}. Inclusion of the effect of instabilities leads to more isotropic attractor labeled ``Instabilities''. (Right) Cartoon of particle propagating through plasma unstable magnetic fields. Each patch of same-sign $B$-field gives rise to a momentum transfer $\Delta p_{\rm kick} \sim  g B l_{\rm coh}$, leading to $\hat{q}\sim (\Delta p_{\rm kick})^2/l_{\rm coh}$. }
\label{fig2}
\end{figure}

\Subsection{ Longitudinal expansion }
The effect of the spatial expansion translates into redshift of the $p_z$ components of the momenta ${\rm d}p_z/{\rm d}t\sim  - p_z/t$ and 
reduces particle number $n'(t)\sim - n(t)/t$.  
If the system is highly anisotropic $d>0$, the energy $E \sim \sqrt{p_z^2 + p_\perp^2}$ of the hard particles is dominated by $p_\perp$, and
the red-shifting does not appreciably affect the particle energies. Then also the energy density scales as $e'(t) \sim -e(t)/t$.
Therefore, at later times energy conservation forces the system to be constrained on other lines of fixed (smaller) energy density on the $cd$-plane, $d=c-1+a$ for $\Qs \tau \sim \alpha^{-a}$, as displayed in the Fig.~\ref{fig2}.
Where exactly on these lines the system takes its place depends on the details of dynamics.

\Subsection{ Momentum broadening and anisotropic screening}
An $\mathcal{O}(1)$ part of momentum transfer experienced  by a hard particle propagating
through an equilibrium plasma comes from momentum exchanges of the order of the screening scale.
In an anisotropic plasma there are magnetic modes at the screening scale which become large,
and hence the soft momentum transfers in an anisotropic system become enhanced and even more important.

In an isotropic medium with $f(\vec{p})\sim f_{\rm iso}(|p|)$, long wavelength chromo-electric fields are stabilized
by the physics of screening: the introduction of a background electric $E$ field deflects trajectories of hard particles in a way
that induces a current $J$; the electric field is reduced by the current, eventually canceling it. Now, the energy 
originally deposited in the electric field is stored in currents, which start to create an 
electric field that in its turn will quench the currents.
Thus, the energy oscillates between a charge separating deformation of the distribution of
hard particles and the electric field. The frequency of the plasma oscillation is related to the screening 
scale $\omega_{\rm pl}^2 |_{f_{\rm iso}} \sim m^2 \equiv \alpha \int {\rm d}^3 p f/p$,
and the dispersion relation of modes with $p \sim m$ becomes $\omega(p)\sim \sqrt{p^2 + \omega_{\rm pl}^2}$.
In an isotropic medium, static magnetic fields are not screened: the deformation to the hard particle distribution due to a 
static $B$-field is simply an overall rotation around the axis set by a magnetic field. The deformed distribution is 
identical to the original one, and therefore no net currents are created, 
and correspondingly the ``magnetic plasma frequency'' $\omega^2_{\rm mag}|_{f_{\rm iso}}\sim 0$; magnetic fields are neither stabilized
nor destabilized (corresponding to $\omega_{\rm mag}^2<0$) by screening. 

In an anisotropic plasma with an angle dependent particle distribution $f_{\rm aniso}(\hat{p})$ the rotation does not leave the hard distribution unchanged
and leads to non-zero currents. Therefore 
the $B$-fields may be stabilized or destabilized. For an anisotropic distribution,
we may nevertheless ask what is the effect averaged over the direction $\hat{B}=\vec{B}/|B|$ (and polarizations) of the magnetic field, $\langle \omega_{\rm mag}^{2}(\hat{B})|_{f_{\rm aniso}(\hat{p})}\rangle_{\hat{B}}$.
This is equivalent to angle averaging over the particles' momentum distributions, and hence corresponds to some isotropic system with $f_{\rm ave}= \langle f(\hat{p})\rangle_{\hat{p}}$, which is neither stabilized nor destabilized
$\omega_{\rm mag}^2(\hat{B})|_{f_{\rm ave}}=0$.
Therefore, even in anisotropic systems, the medium's impact on magnetic fields, averaged over all 
directions, is neutral.%
\footnote{
In terms of HL effective theory, this is to say that the angle average of the HL gluon polarization tensor is proportional to the isotropic, thermal 
polarization tensor.
}
 However, if there are any directions that are stabilized ($\omega_{\rm mag}^2>0$), there must be other
directions that have the opposite effect and are destabilized ($\omega_{\rm mag}^2<0$).
These are the plasma unstable modes. They are \emph{always} present in anisotropic systems,
and they grow exponentially $B(t)\sim B(0) e^{\gamma t}$ with a growth rate $\gamma\sim m$.

Which modes become unstable depends on the details of the anisotropic distribution. The 
set of unstable modes can be found, \emph{e.g.}, by finding the range of momenta for 
which the retarded HL propagator has poles in the upper half complex-plane. This question has been addressed in \cite{RS1,AM}, and for the distribution
of Eq.~(\ref{dist}), the unstable modes have $k^{\rm inst}_\perp\lesssim m$ and $k^{\rm inst}_z \sim \alpha^{-d} m$.

Nothing grows exponentially forever, and the growth of the magnetic fields 
stops when some new physics kicks in. The previous discussion relied on linearizing in
the size of the magnetic field; once dynamics become non-linear, the non-perturbative
interactions stop the exponential growth and the instabilities saturate.  The non-linear 
physics enters through the covariant derivative $D_\mu = \partial_\mu + i g A_\mu$, 
and therefore the non-linear physics kicks in when the 
interaction term competes with the derivative in $D_\mu$. In a non-abelian theory, the 
magnetic modes can interact among themselves, so the relevant momentum scale is 
that of the instabilities $\partial_\mu \sim k^{\rm inst}_{\perp}$, corresponding to $A \sim k_{\perp}^{\rm inst}/g$, or 
$B^2\sim (\Delta \times A)^2 \sim (k^{\rm inst} A)^2 \sim {k_\perp^{\rm inst}}^2 k_{\rm inst}^2/\alpha$, or equivalently $f\sim 1/\alpha$.%
\footnote{
In \cite{KM1}, a more precise gauge invariant criterion is presented in terms of Wilson loops. For here, this simpler criterion is however good enough.
}
That the non-linear interaction indeed saturates the instability has been observed in numerical
HL simulations \cite{fate}.

How important for dynamics the unstable magnetic fields are depends on how large they 
grow and how strongly correlated they are. When the hard particles move through the 
large magnetic fields, they experience a time varying Lorentz force (see Fig.~\ref{fig2}). The magnetic field
remains coherently in the same direction for the characteristic coherence length of 
the magnetic field, $l_{\rm coh}\sim {k^{\rm inst}_\perp}^{-1}$ for a particle moving in a $\perp$ direction. The momentum accumulated 
in one coherence length is $\Delta p_{\rm kick} \sim g B l_{\rm coh}$, and propagating
through many such uncorrelated patches of coherent magnetic field leads to diffusive
momentum transfer described by a momentum diffusion coefficient 
$\hat{q}_{\rm inst}\sim (\Delta p_{\rm kick})^2/l_{\rm coh}\sim \alpha B^2/k_{\perp}^{\rm inst}
\sim k_{\rm inst}^2 k^{\rm inst}_\perp\sim \alpha^{-2 d}m^3\sim \alpha^{-\frac{1}{2}d+\frac{3}{2}(1-c)}Q_s^3$. As expected, this plasma instability induced
$\hat{q}$ is larger than the one due to perturbative elastic scattering $\hat{q}_{\rm el}\sim \alpha^{-2c+d}Q_s^3$ during the equilibration process, 
and therefore we do not need to discuss elastic scattering in the following.%

\Subsection{Competition between momentum broadening and longitudinal expansion}
We are now ready to solve the trajectory of the system in the $cd$-plane during the first and second stages.
The longitudinal expansion makes the distribution more anisotropic with a rate $\d p_{z} / \d \tau = -p_z/\tau $
corresponding to $\d d / \d a|_{\rm expansion} = + 1$, while the momentum broadening isotropizes the distribution with 
a rate $\d p_z/\d \tau \sim \hat{q}_{\rm inst}/p_z$ corresponding to $\d d / \d a |_{\rm inst} \sim -\alpha^{-\frac{5}{2}d+\frac{3}{2}(1-c)-a} \sim +\alpha^{-4d + \frac{1}{2} a}$,
with the energy conservation condition $1-c = a-d$. 
These two opposite effects can compete only if also $\d d / \d a|_{\rm inst} \sim \mathcal{O}(\alpha^{0})$, which 
happens when $d_{\rm att.}(a)= \frac{a}{8}$, and $c_{\rm att.} =1-\frac{7}{8} a$.
This is solution plotted  on the $cd$-plane in Fig.~\ref{fig2}.%
\footnote{We also find another, weakly anisotropic attractor. For details, and why we do not think it is realized, see \cite{KM1}.}

The solution is attractive in the sense that if the system is in a state below (above) the attractor ($d < \frac{a}{8}$), the 
system becomes more anisotropic (isotropic) and reaches the attractor along a line of constant energy density in a time scale comparable to 
the age of the system. 

In Fig.~\ref{fig2} we also plot the corresponding attractor neglecting the effect of plasma instabilities, so that 
the momentum broadening is due ordinary elastic scattering. As the interaction is weaker the distribution is,
at a given time, less isotropic than with instabilities. The attractor has a
corner at $f\sim1$ where the system turns from over- to under-occupied and $\hat{q}_{\rm el} \propto f(1+f)$ changes behavior. For a plasma instability driven
system, there is no such kink in the attractor; the plasma instabilities depend on the hard particle distribution
only through the scale $m$, which is linear in $f$, and the attractor is valid also at $\Qs \tau\gg \alpha^{-\frac{8}{7}}$.
If no new physics would kick in, the system would not thermalize and would only get more and more dilute
as a function of time.

\Section{Creation of a soft nearly thermal bath, ``Bottom-up'', $\alpha^{-\frac{12}{5}} \ll \Qs\tau\ll \alpha^{-\frac{5}{2}}$}
\label{sec:bottomup}
\begin{figure}
\vspace{-0.5cm}
\begin{center}
\includegraphics[height=0.16\textheight]{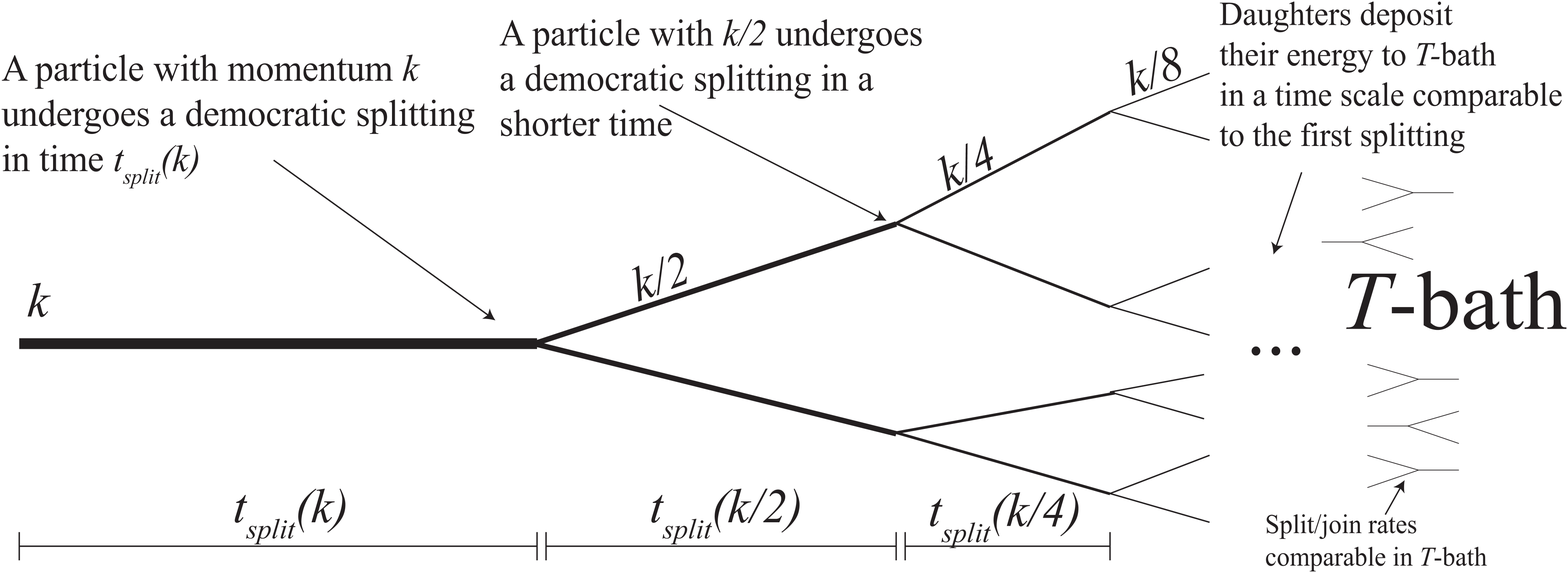}
\hspace{0.92cm}
\includegraphics[width=0.28\textwidth]{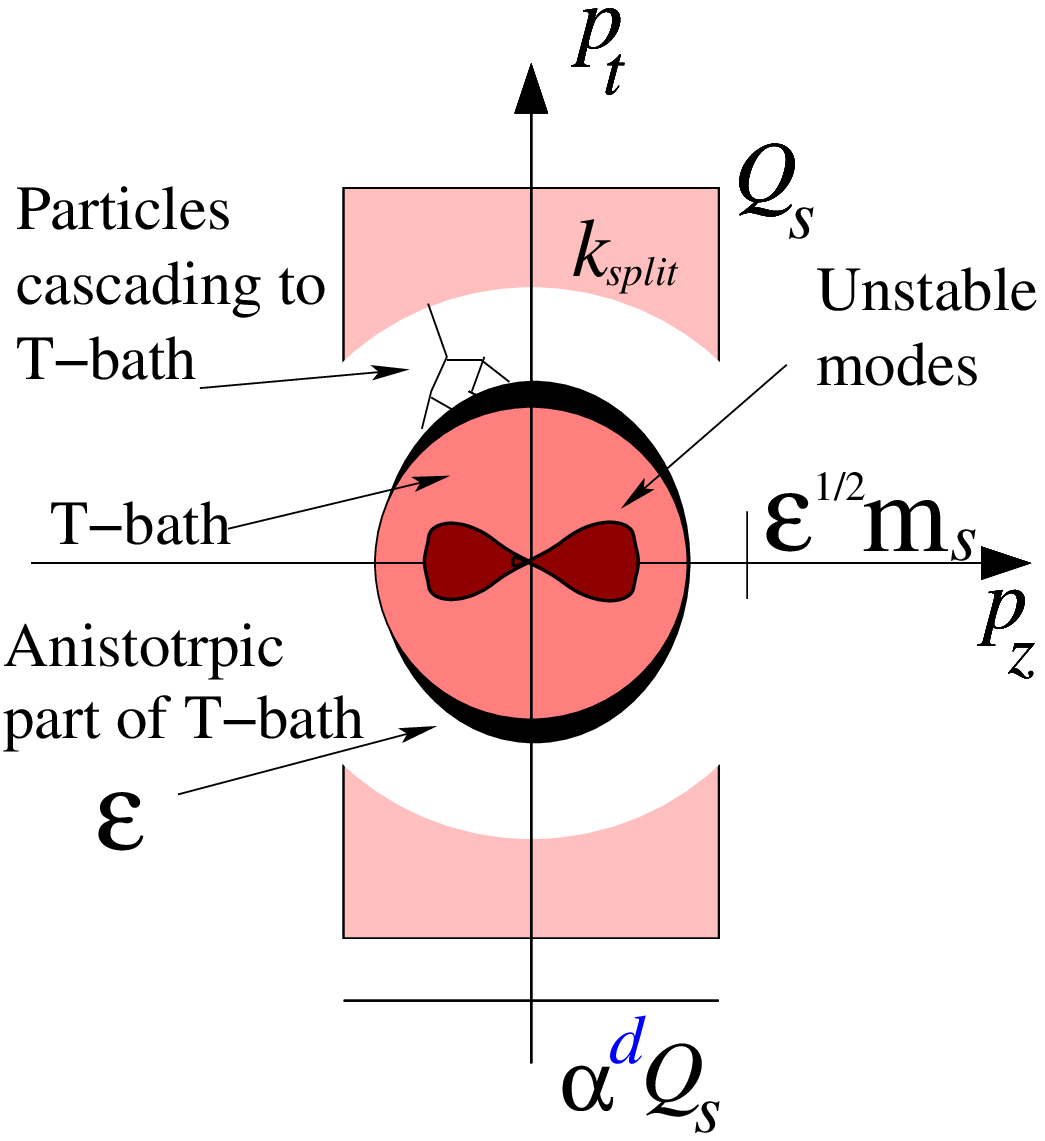}
\end{center}
\vspace{-0.4cm}
\caption{
(Left) Cartoon of the infrared cascade of a particle with momentum $k$. Once the particle has had time to split in two daughters with 
comparable momenta, the daughters quickly cascade to the infrared until they reach the soft thermal bath. (Right) Cartoon of the momentum scales in late stages of the 
evolution. The important scales are (from hard to soft): highly anisotropic distribution of hard particles at scale $\Qs$ (pink), scale $k_{\rm split}$ below which particles have had time to cascade to infrared, soft nearly thermal bath with anisotropy $\epsilon$ (black and pink), and the anisotropic screening scale  of the $T$-bath $\epsilon^{1/2}m$ with strong magnetic fields (dark red). The system becomes effectively thermalized once $k_{\rm split}$ reaches $\Qs$.}
\label{fig3}
\end{figure}

During the first stage, inelastic scattering in an over-occupied system works towards joining 
the hard particles and cascade energy to the ultraviolet. Once the system becomes under-occupied,
this behavior changes qualitatively and the inelastic scattering starts to \emph{split} hard particles and 
cascade their energy to the infrared. 
This leads to the creation of a two-scale system; in addition to the original population
of hard particles, the debris of splitting form a new, \emph{soft}%
\footnote{Soft compared to $\Qs$ but still hard compared to the screening scale}, population. The constituents of the soft bath
have less inertia, and hence they may isotropize and thermalize faster 
(by the time $\Qs \tau \sim \alpha^{-\frac{56}{25}}$, see \cite{KM1} for details). 

How much energy is deposited in the soft bath and what its temperature is depend on how
effectively the hard particles cascade to the infrared. 
The rate at which a hard particle undergoing transverse momentum diffusion emits daughters 
of momentum $p_{\rm daught.}$ is given (in the LPM regime) by 
$t_{\rm split}^{-1}\sim \alpha \sqrt{\hat{q}/p_{\rm daught.}}$. That is, by the time $\tau$
all particles with momentum $k < k_{\rm split}\sim \alpha^2 \hat{q} \tau^2$ have had enough time to emit a daughter
whose energy is comparable to the emitter's, corresponding to splitting the original particle democratically into two daughters with 
half the original energy (see Fig.~\ref{fig3}).  The daughters have a higher splitting rate than the mother particle, and they undergo successive
democratic splittings in a time that is shorter than the age of the system, cascading their energy
to the soft bath.  As by the time $\tau$ each hard particle have had time to emit $\mathcal{O}(1)$ particle with energy $k_{\rm split}$, 
the energy density and the temperature of the soft bath is $e \sim T_{\rm soft}^4 \sim k_{\rm split} \int d^3p f(p)$.%
\footnote{For $\Qs \tau \ll \alpha^{-\frac{12}{5}}$, $\hat{q}$ is a strongly angle dependent function
giving rise to subtleties not relevant for discussion here.}

Even if the soft sector does not dominate the energy density ($\int d^3p f(p) p$),
it can dominate other characteristics of the medium, such as screening ($\int d^3p f(p)/p$) and $\hat{q}$.  
When $\hat{q}$ is dominated by interactions with the
soft bath, the hard particles effectively decouple from each other and see only the soft medium. In this case the physical picture is 
that of a system consisting of a nearly thermal bath through which a distribution of few but highly energetic ``jets'' with energy $\Qs$ propagate. The interaction
with the medium quenches the jets via radiative energy loss. When the jets have had time to lose all their energy to 
the medium, that is $k_{\rm split}\sim \Qs$ or equivalently $\tau_{\rm eq}\sim \alpha^{-1}\sqrt{\Qs/\hat{q}}$, the system has effectively thermalized. 

How fast the system then thermalizes is then controlled by how strongly the soft nearly thermal 
bath broadens the momentum of the jets. In \cite{BMSS}, it was assumed that the primary mechanism
is through perturbative elastic scattering, so that $\hat{q}_{\rm el} \sim\alpha^2 T_{\rm soft}^3$. At the time scale the hard jets 
deposit their energy to the thermal bath, the energy density of the thermal bath is comparable to that
of the hard jets, $T_{\rm soft}^4\sim \alpha^{-1}\Qs^4/(\Qs\tau)$, leading to an estimate for the thermalization time $\Qs\tau_{0}\sim \alpha^{-13/5}$.

However, also in this case the plasma instabilities dominate the momentum transfer. The soft nearly thermal bath
is also anisotropic, partly due to the ``stretching'' by expansion and partly because the cascade of the hard particles heating the bath
is anisotropic. Both of these effects lead to a parametrically weak anisotropy of the soft bath, of order $\epsilon \equiv  \langle |p_{\perp}|\rangle / \langle |p_z| \rangle \sim T^2/(\hat{q} \tau)$. In \cite{KM1}, we found that weakly anisotropic systems such as this generates unstable modes 
in a range $k^{\rm inst}_z \sim k^{\rm inst}_\perp \sim \epsilon^{1/2} m_s $, where $m_s^2 \sim \alpha T^2$ is the screening scale of the soft distribution.%
\footnote{
In fact the $k^{\rm inst}$ is given by the screening scale of the anisotropic part of the distribution $m^2_{\rm aniso}\sim \epsilon\alpha T^2 $ as the 
isotropic part does not stabilize nor destabilize the modes.
}
These unstable modes then subsequently cause momentum broadening $\hat{q}_{\epsilon}\sim k_{\rm inst}^3\sim \epsilon^{3/2} \alpha^{3/2}T^3$,
which is larger than the elastic contribution arising from the soft sector as long as $\epsilon \gg \alpha^{1/3}$, and dominates over $\hat{q}$
arising from the hard distribution for $\Qs \tau \gg \alpha^{-12/5}$. See Fig~\ref{fig3} for cartoon of the scales.

Taking everything together and solving self-consistently gives
\begin{equation}
\begin{array}{rclrclrclrcl}
k_{\rm split} &\sim& \alpha^2 \hat{q}_{\rm \epsilon} \tau^2,&
T^4   &\sim& k_{\rm split} \Qs^3 ( \alpha \Qs \tau)^{-1}, & 
\hat{q}_\epsilon &\sim& \epsilon^{3/2}\alpha^{3/2}T^3, &
\epsilon &\sim& T^2/(\hat{q}_\epsilon\tau)\\
k_{\rm split}&\sim& \alpha^5 \Qs  (\Qs \tau)^2, &
T &\sim& \alpha \Qs (\Qs \tau)^{1/4}, &
\hat{q}_{\epsilon}&\sim& \alpha^3 \Qs^3, &
\epsilon &\sim& \alpha^{-1}(\Qs \tau)^{-1/2},
\end{array}
\end{equation}
so that the hard particles have had time to cascade to the thermal bath, \emph{i.e.}  $k_{\rm split }\sim\Qs$, by the time $\Qs \tau \sim \alpha^{-5/2}$. At this point the system (especially its $T_{\mu \nu}$) is approximately isotropic and is amendable to a hydrodynamical description.


\vspace{-0.3cm}
\begin{thebibliography}{99}
\vspace{-0.3cm}
\bibitem{KM1}
  A.~Kurkela and G.~D.~Moore,
  JHEP {\bf 1112} (2011) 044
  [arXiv:1107.5050 [hep-ph]];
  A.~Kurkela and G.~D.~Moore,
  JHEP {\bf 1111} (2011) 120
  [arXiv:1108.4684 [hep-ph]].
  \bibitem{CGC}
  E.~Iancu and R.~Venugopalan,
  In *Hwa, R.C. (ed.) et al.: Quark gluon plasma* 249-3363
  [hep-ph/0303204].
  \bibitem{glasmamelt} 
  P.~Romatschke and R.~Venugopalan,
  Phys.\ Rev.\ Lett.\  {\bf 96} (2006) 062302
  [hep-ph/0510121].

  
\bibitem{BMSS}
  R.~Baier, A.~H.~Mueller, D.~Schiff and D.~T.~Son,
  Phys.\ Lett.\ B {\bf 502} (2001) 51
  [hep-ph/0009237].
  
  \bibitem{instab}
  S.~Mrowczynski,
  Phys.\ Lett.\ B {\bf 214} (1988) 587
   [Erratum-ibid.\ B {\bf 656} (2007) 273];
 
  
    \bibitem{clas}
  J.~Berges and S.~Schlichting,
  Phys.\ Rev.\ D {\bf 87} (2013) 014026
  [arXiv:1209.0817 [hep-ph]].
  
  
 \bibitem{kinetic}
  P.~B.~Arnold, G.~D.~Moore and L.~G.~Yaffe,
  JHEP {\bf 0301} (2003) 030
  [hep-ph/0209353].

\bibitem{BE1}
  J.~-P.~Blaizot, F.~Gelis, J.~-F.~Liao, L.~McLerran and R.~Venugopalan,
  Nucl.\ Phys.\ A {\bf 873} (2012) 68
  [arXiv:1107.5296 [hep-ph]].

  \bibitem{KM3}
    A.~Kurkela and G.~D.~Moore,
  Phys.\ Rev.\ D {\bf 86} (2012) 056008
  [arXiv:1207.1663 [hep-ph]].


  \bibitem{RS1}
  P.~Romatschke and M.~Strickland,
  Phys.\ Rev.\ D {\bf 68} (2003) 036004
  [hep-ph/0304092].
  
  \bibitem{AM}
  P.~B.~Arnold and G.~D.~Moore,
  Phys.\ Rev.\ D {\bf 76} (2007) 045009
  [arXiv:0706.0490 [hep-ph]].
    
\bibitem{fate}
  P.~B.~Arnold, G.~D.~Moore and L.~G.~Yaffe,
  Phys.\ Rev.\ D {\bf 72} (2005) 054003
  [hep-ph/0505212].

\end{thebibliography}
\end{document}